\documentclass[%
aps,
reprint,
superscriptaddress,
nofootinbib,
nobibnotes,
showkeys,
floatfix,
]{revtex4-2}

\usepackage{xcolor}
\usepackage{graphicx}
\graphicspath{{./fig/}}
\usepackage{amsmath,amssymb,amsfonts,mathrsfs,bm}
\usepackage[%
colorlinks=true,
linkcolor=blue,
citecolor=blue,
]{hyperref}

\newcommand{\dif}{\mathrm{d}}

\newcommand{\Msun}{M_{\odot}}
\newcommand{\yr}{\mathrm{yr}}
\newcommand{\eV}{\mathrm{eV}}
\newcommand{\pc}{\mathrm{pc}}

\newcommand{\Gpc}{\mathrm{Gpc}}

\newcommand{\Hz}{\mathrm{Hz}}

\newcommand{\DM}{\mathrm{DM}}
\newcommand{\SF}{\mathscr{S}}
\newcommand{\CF}{\mathscr{C}}
\newcommand{\varphip}{\varphi_{\mathrm{p}}}
\newcommand{\varphit}{\tilde{\varphi}}
\newcommand{\FDM}{F_{\mathrm{DM}}}
\newcommand{\rhoDM}{\rho_{\mathrm{DM}}}
\newcommand{\rhoDMb}{\bar{\rho}_{\mathrm{DM}}}
\newcommand{\Mc}{M_{c}}
\newcommand{\dL}{d_{L}}
\newcommand{\Rx}{R_{*}}
\newcommand{\tp}{t_{\mathrm{p}}}
\newcommand{\ttld}{\tilde{t}}

\newcommand{\lh}{\hat{\lambda}}
\newcommand{\ret}{\mathrm{ret}}

\newcommand{\Gyr}{\,\mathrm{Gyr}}

\newcommand{\mueV}{\,\mu\mathrm{eV}}
\newcommand{\GeV}{\,\mathrm{GeV}}
\newcommand{\osc}{\mathrm{osc}}
\newcommand{\MW}{\mathrm{MW}}
\newcommand{\mc}{\mathrm{mc}}
\newcommand{\hmin}{\mathrm{h,min}}

\begin{document}

\title{Gravitational Wave Duet by Resonating Binary Black Holes within Ultralight Dark Matter
}

\author{Jeong Han Kim}
\email{jeonghan.kim@cbu.ac.kr}
\affiliation{Department of Physics, Chungbuk National University, Cheongju, Chungbuk 28644, Korea}

\author{Xing-Yu Yang}
\email[Corresponding author:~]{xingyuyang@kias.re.kr}
\affiliation{Quantum Universe Center (QUC), Korea Institute for Advanced Study, Seoul 02455, Republic of Korea}

\begin{abstract}
    Gravitational wave observations have significantly broadened our capacity to explore fundamental physics beyond the Standard Model, providing crucial insights into dark matter that are inaccessible through conventional methods. Here, we investigate the resonant interactions between binary black hole systems and solitons, self-gravitating configurations of ultralight bosonic dark matter, which induce metric perturbations and generate distinct oscillatory patterns in gravitational waves. Upcoming experiments such as the Laser Interferometer Space Antenna could detect the oscillatory patterns in gravitational waveforms, providing an evidence for solitons. Because the effect relies solely on gravity, it does not assume any coupling of the dark sector to Standard Model particles, highlighting the capability of future gravitational-wave surveys to probe dark matter.
\end{abstract}

\maketitle

\section{Introduction}
\label{sec:Intro}
After a long period of extensive efforts to unravel the nature of dark matter (DM), including direct and indirect searches \cite{Graham:2015ouw, Schumann:2019eaa, Gaskins:2016cha} as well as collider studies \cite{Boveia:2018yeb}, the precise properties of DM, such as its mass and self-coupling strength, remain elusive.
This ongoing pursuit stands as one of the most colossal puzzles in modern cosmology.
A key question moving forward is how we will address this challenge and advance our understanding of DM lurking in the structure of the universe. The lack of definitive results has opened new avenues in the search for DM, leading to diverse approaches.

Gravitational wave (GW) observations have greatly expanded our ability to investigate DM. Specifically, the GW signals emitted by binary black holes offer new opportunities to probe DM surrounding these massive objects. In this study, we focus on one such candidate for ultralight DM (ULDM)~\cite{Ferreira:2020fam,Hui:2021tkt}%
\footnote{Throughout this work we remain agnostic about the specific ultraviolet origin of ULDM, although axion-like particles (ALPs) are often motivated by Peccei-Quinn mechanism~\cite{Peccei:1977hh, Peccei:1977ur, Weinberg:1977ma, Wilczek:1977pj, Demirtas:2018akl, Halverson:2017deq, Cicoli:2012sz, Acharya:2010zx, Arvanitaki:2009fg, Choi:2009jt, Conlon:2006tq, Svrcek:2006yi, Witten:1984dg}.}: a bosonic DM field characterized solely by its mass $m$ with self-coupling $\lambda$.
Prior studies have indicated that GWs could potentially reveal these particles through a black hole superradiance \cite{Arvanitaki:2010sy, Arvanitaki:2014wva, Arvanitaki:2016qwi, Boskovic:2024fga}, axion-mediated forces and radiations in binary neutron stars \cite{Hook:2017psm, Huang:2018pbu},
scalar field emissions caused by its interaction with gravity \cite{Maselli:2021men},
frequency modulations~\cite{Wang:2023phr} and phase deviations~\cite{Brax:2024yqh} as GWs transverse oscillating scalar fields,
dynamical frictions \cite{Macedo:2013qea, Kim:2022mdj, Kadota:2023wlm, Boudon:2023vzl, Kim:2022mdj}, and
deviations in the speed of GWs \cite{Dev:2016hxv}.

\begin{figure}[t!]
    \includegraphics[width=\columnwidth]{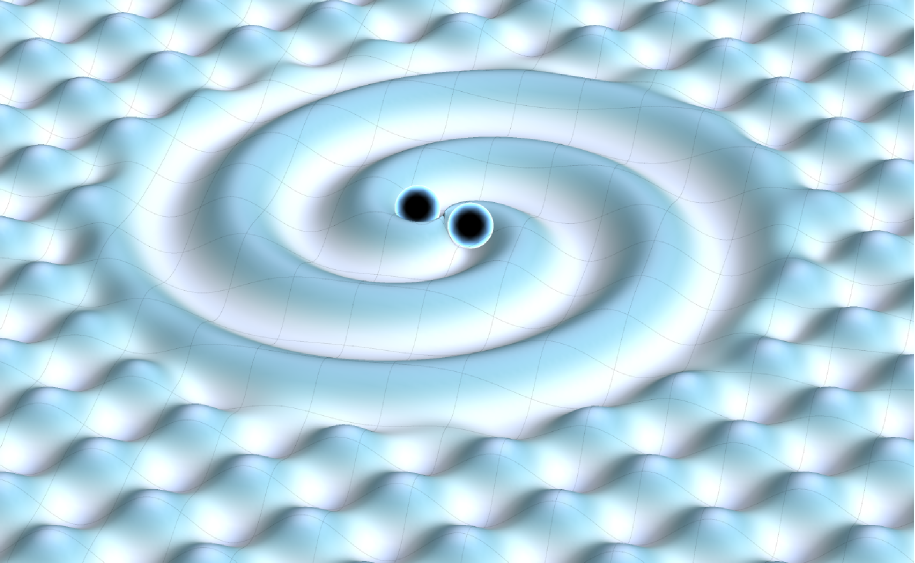}
    \caption{\textbf{Illustration of binary black holes resonating with ULDM.} The spacetime perturbations induced by ULDM introduce an additional force between the black holes, leading to distinctive patterns in the GWs emitted during their merger.}
    \label{fig:illustration}
\end{figure}

The wave-like nature of ULDM can induce oscillatory metric perturbations, as demonstrated in previous studies~\cite{Blas:2019hxz, Blas:2016ddr, Boskovic:2018rub, Rozner:2019gba, Desjacques:2020fdi}. In this work, for the first time, we investigate the resonant interactions between binary black hole systems and ULDM (see Fig.~\ref{fig:illustration} for an illustration).
As the binary black hole system loses energy and spirals inward, its growing orbital frequency periodically oscillates with the oscillation frequencies of ULDM.
These instances of resonance alter the gravitational waveforms, as the frequency of binary black holes sweeps through various harmonics.
Such variations appear as distinct patterns in the waveform, particularly alterations in the frequency and phase of the GWs.
These changes can be identified by observatories such as LISA~\cite{LISA:2017pwj}, Taiji~\cite{Ruan:2018tsw}, Tianqin~\cite{TianQin:2015yph}, and Deci-hertz Interferometer Gravitational Wave Observatory (DECIGO) \cite{Seto:2001qf}.
By using a detailed Fisher matrix analysis, we will show that our approach enables exploration across an extensive area of the parameter space for $m$ and $\lambda$.

Our paper is structured as follows. Section~\ref{sec:ULDM} introduces a simplified model of ULDM, and examines how they can cause oscillations in the spacetime metric. Section~\ref{sec:BBH} then describes the evolution of a binary black hole system under the influence of oscillating ULDM, explaining how these interactions alter black hole dynamics. Section~\ref{sec:GWs} presents our main results from the analysis of GW signals, shedding light on the ULDM parameter space which future experiments such as LISA can probe.
Finally, Section~\ref{sec:dis} concludes with a discussion on the implications of our findings and potential directions for future research.

\section{Oscillations in Spacetime Induced by ULDM}
\label{sec:ULDM}
We study a generic real scalar field $\phi$ charged under a $\mathbb{Z}_2$ symmetry, whose dynamics are governed by the action
\begin{equation}
    S=\int \dif^{4}x \sqrt{-g} \left[ \frac{M^2_{\rm pl}}{2}R -\frac{1}{2} g^{\mu\nu} \partial_{\mu}\phi \partial_{\nu}\phi - V(\phi) \right],
    \label{eq:action}
\end{equation}
where $M_{\rm pl} = \sqrt{1/8 \pi G}$ denotes a reduced Planck mass%
\footnote{Unless otherwise stated, we use natural units $c=\hbar=1$ for the notational brevity.}, $R$ is the Ricci scalar, $V(\phi) = m^2\phi^2/2 + \lambda \phi^4/4!$ represents the potential of the scalar field with its mass $m$ and a quartic self-coupling constant $\lambda$.
We assume a negative coupling $\lambda<0$, which renders the ULDM interaction attractive.
We consider the metric perturbation in Newtonian gauge
\begin{equation}
    ds^2 = -[1 + 2 \Phi (t, \bm{x}) ]dt^2 + [1 - 2 \Psi(t, \bm{x}) ] \delta_{ij} dx^i dx^j .
    \label{eq:metric}
\end{equation}
Here, $\Phi(t, \bm{x})$ and $\Psi(t, \bm{x})$ represent scalar perturbations in spacetime.

ULDM can form a stable soliton made of condensed fields with random phases, producing an interference pattern of wave packets, each approximately the size of the de Broglie wavelength, $\lambda_{\text{dB}} = 2 \pi / (m v)$, where $v$ is a virial velocity.
Within each wave packet, the fields exhibit coherent oscillations, and ULDM behave like a single macroscopic fluid in the non-relativistic limit~\cite{Boudon:2022dxi, Brax:2019fzb}
\begin{equation}
    \phi (t, \bm{x}) = \phi_0 (\bm{x}) \cos (\omega_a t + \Upsilon(\bm{x})) \;,
    \label{eq:ansatz}
\end{equation}
where $\phi_0 (\bm{x})$ and $\Upsilon(\bm{x})$ are functions that exhibit slow changes in positions, and $\omega_a = m (1+ \lambda \phi^2_0/(16 m^2))$ denotes the angular frequency of ULDM. While the self-gravity of ULDM contributes to their angular frequency, our main focus is on how their self-interaction impacts this frequency. To simplify the discussion, we will neglect the contribution from self-gravity.

The corresponding energy-momentum tensor reads
\begin{equation}
    T^{\mu}_{\;\;\nu} = g^{\mu \alpha} \partial_{\alpha}\phi \partial_{\nu} \phi
    - \delta^{\mu}_{\;\;\nu} \big(\frac{1}{2} g^{\rho \sigma} \partial_{\rho} \phi \partial_{\sigma} \phi + V(\phi) \big) \;.
    \label{eq:Tmunu}
\end{equation}
The energy density for the ULDM can be derived from the time-time component of the energy-momentum tensor.
Averaging the energy density of ULDM, specifically the time-time component of the energy-momentum tensor, over one period of density oscillation, $2 \pi / \omega_{a}$, yields%
\footnote{Note that the density profile of ULDM varies with the distance from the center. A detailed examination of this density profile requires solving the Schr\"{o}dinger-Poisson equations to accurately determine the distribution of ULDM, which we intend to address in future work.}
\begin{equation}
    \rhoDMb = \frac{1}{2} m^2 \phi_{0}^{2} + \frac{3\lambda}{64}\phi_{0}^{4} + \frac{\lambda^2}{1024 m^2}\phi_{0}^{6} \;,
    \label{eq:rhoDM}
\end{equation}
where we have neglected the terms with spatial derivatives in $\phi$.
However, solitons cannot sustain densities beyond a certain threshold, as they become unstable and collapse (detailed stability criteria are provided in Appendix \ref{sec:stability}).
In our study, we have selected benchmark values for $\rhoDMb$ ranging from $10^{16}$ to $10^{20} \Msun / \pc^3$, which are compatible with the stability criteria. Such high densities can, for example, be realized in the axion minicluster framework~\cite{Hogan:1988mp, Kolb:1993zz, Kolb:1993hw, Kolb:1994fi, Kolb:1995bu, Zurek:2006sy, Hardy:2016mns, Enander:2017ogx}, and we discuss the details in the in Appendix~\ref{sec:Mini}.

Using Eq. (\ref{eq:rhoDM}), the amplitude can be expressed as $\phi_{0}^{2} = \Lambda \rhoDMb / m^2$, where $\Lambda$ is a dimensionless function of $\lh$, and $\lh \equiv \lambda \rhoDMb/m^{4}$ represents the dimensionless self-coupling constant, defined in the range $\lh \in [-\sqrt{64/27},0)$.
See Appendix~\ref{sec:Pressure} for the detailed derivations.

The pressure can be computed from the spatial components of the the energy-momentum tensor from Eq.(\ref{eq:Tmunu})
\begin{equation}
    \begin{aligned}
        P_{\DM} = -\rhoDMb \Big[ \Lambda_{0}
 &+ \Lambda_{2} \cos(2 \omega_a t+2\Upsilon) \\
 &+ \Lambda_{4} \cos(4 \omega_a t+4\Upsilon) \Big] \;,
    \end{aligned}
    \label{eq:pDM}
\end{equation}
where
\begin{subequations}
    \begin{align}
        \Lambda_{0} =& -\frac{\lh}{64}\Lambda^{2} - \frac{\lh^2}{1024}\Lambda^{3} \;, \\
        \Lambda_{2} =& \frac{1}{2} \Lambda + \frac{5\lh}{96}\Lambda^{2} + \frac{\lh^2}{1024 }\Lambda^{3} \;, \\
        \Lambda_{4} =& \frac{\lh}{192}\Lambda^{2} \;.
    \end{align}
    \label{eq:lambdas}
\end{subequations}
Thus, the ULDM pressure consists of a constant part $\Lambda_{0}$, along with time-varying terms $\Lambda_{2}$ and $\Lambda_{4}$ with oscillating at frequencies of $2\omega_a$ and $4\omega_a$ respectively.
These dimensionless parameters are confined within the range $-1 \leq \Lambda_{0,2,4} \leq 1$.
It should be noted that the harmonic component does not require a non-zero self-interaction.
Even when $\lh \rightarrow 0$, the amplitude $\Lambda_{2}$ remains non-zero (see Appendix~\ref{sec:Pressure} for more details).

To study oscillations induced by ULDM in spacetime, we calculate perturbed Einstein equations
\begin{align}
    \nabla^2 \Psi &= 4 \pi G \rhoDM \;, \label{eq:Ein00} \\
    \ddot{\Psi} + \frac{1}{3} \nabla^2 (\Phi - \Psi) &= 4 \pi G P_{\DM} \;, \label{eq:Einij}
\end{align}
Neglecting the spatial gradients and using Eq.(\ref{eq:Einij}), we obtain (see Appendix~\ref{sec:Eins} for more details)
\begin{equation}
    \ddot{\Psi} = - 4 \pi G \rhoDMb \Big[ \Lambda_2 \cos(2 \omega_a t+2\Upsilon) + \Lambda_4 \cos(4 \omega_a t+4\Upsilon) \Big] \;.
    \label{eq:psidd2}
\end{equation}
This equation describes the change in the metric perturbation over time at frequencies $2\omega_a$ and $4\omega_a$, influenced by the ULDM pressure.

\section{Binary black holes}
\label{sec:BBH}
We consider a binary system consisting of black holes with masses $M_1$ and $M_2$, surrounded by a cloud of ULDM.
The metric perturbation induced by ULDM can generate an additional force between the black holes. The Fermi normal coordinates provide a convenient way to express the geodesic deviation equations for the binary~\cite{1978ApJ, Blas:2019hxz}
\begin{equation}
    \ddot{\bm{r}} = -\FDM \hat{\bm{r}} \;,
    \label{eq:ddr}
\end{equation}
where $\bm{r}$ is a vector connecting the two bodies, $\hat{\bm{r}}$ is the corresponding unit vector, and $\FDM=\ddot{\Psi}r$ denotes the exerted force.
In our analysis, at a separation distance of approximately $\mathcal{O}(100)$ Schwarzschild radii between the binary black holes, we assume that the ULDM density is locally homogeneous and isotropic relative to the barycenter of the binary system, leading to a radially exerted force.
This additional force perturbs the Keplerian orbit of the binary system~\cite{poisson2014gravity}
\begin{subequations}
    \begin{align}
        \frac{\dif a}{\dif t} =& -2\sqrt{\frac{a^{3}}{GM}}\frac{e}{\sqrt{1-e^{2}}}\sin(\varphi - \varphip) \FDM \;, \label{eq:orbital1}\\
        \frac{\dif e}{\dif t} =& -\sqrt{\frac{a}{GM}}\sqrt{1-e^{2}}\sin(\varphi - \varphip)\FDM \;, \label{eq:orbital2}\\
        \frac{\dif \varphip}{\dif t} =&\; \sqrt{\frac{a}{GM}}\frac{\sqrt{1-e^{2}}}{e}\cos(\varphi - \varphip)\FDM \;, \label{eq:orbital3} \\
        \frac{\dif \varphi}{\dif t} =&\; \sqrt{\frac{GM}{a^{3}}}\frac{[ 1+e\cos(\varphi-\varphip) ]^{2}}{(1-e^{2})^{3/2}} \;, \label{eq:orbital4}
    \end{align}
    \label{eq:orbital}
\end{subequations}
where $M \equiv M_1+M_2$ is the total mass of the binary, $a$ stands for the semimajor axis, $e$ denotes the eccentricity, $\varphi$ is the orbital angle, and $\varphip$ is the longitude of the pericenter.
Eq.(\ref{eq:orbital1}) demonstrates that a non-zero value of $e$ increases the effect of the additional force on the rate of change in $a$. Similarly, Eq.(\ref{eq:orbital2}) shows that the rate of change in $e$ is influenced by this external force.

In addition, the emission of GWs results in the loss of energy and angular momentum of the binary~\cite{maggiore2008gravitational}
\begin{subequations}
    \begin{align}
        \left\langle \frac{\dif a}{\dif t} \right\rangle=&-\frac{64 G^{3} \mu M^{2}}{5 c^{5} a^{3} (1-e^{2})^{7/2}}\left( 1+\frac{73}{24}e^{2}+\frac{37}{96}e^{4} \right) \;, \\
        \left\langle \frac{\dif e}{\dif t} \right\rangle=&-\frac{304 G^{3}\mu M^{2} e}{15 c^{5} a^{4} (1-e^{2})^{5/2}}\left( 1+\frac{121}{304}e^{2} \right) \;,
    \end{align}
    \label{eq:orbitalGW}
\end{subequations}
where $\mu \equiv M_1 M_2 /M$ is a reduced mass, and $\langle\cdots\rangle$ denotes the average taken over an orbital period, $T=2\pi/\omega$.
When stellar black holes move through a DM environment, they would typically experience dynamical friction. Nontheless, to focus on the main aspect of resonant effects, we will omit the impact of dynamical friction and halo feedback from our discussion, saving it for future work.

To solve the orbital evolution as described in Eqs.(\ref{eq:orbital}-\ref{eq:orbitalGW}), it is convenient to define the following dimensionless quantities
\begin{equation}
    \begin{aligned}
        \alpha \equiv \frac{a}{\Rx}, \;\;
        \tau \equiv \frac{t c}{\Rx}, \;\;
        \eta \equiv \frac{\mu}{M}, \;\; \\
        \zeta \equiv \frac{4 \pi G \rhoDMb \Rx^2}{c^2}, \;\;
        \nu \equiv \Omega/\omega,
    \end{aligned}
\end{equation}
where $\Rx\equiv GM/c^{2}$ represents half of the Schwarzschild radius of a binary, $\omega = \sqrt{GM/a^{3}}$ denotes an orbital frequency of a binary system, and $\Omega = 2 \omega_a $ is related to the oscillation frequency of ULDM.

Applying Fourier decomposition to the orbital elements and averaging these values over the orbital period the orbital equations simplify to (refer to Appendix~\ref{sec:FDorbit} for detailed explanations)
\begin{subequations}
    \begin{align}
        \left\langle \frac{\dif\alpha}{\dif\tau} \right\rangle=
&\zeta \alpha^{5/2} \frac{2 e}{\sqrt{1-e^{2}}} \Big[\Lambda_{2} \sin(\pi \nu + \gamma) \SF(\nu,e) \notag \\
&~~~~~~~~~~~~+ \Lambda_{4} \sin(2\pi \nu + 2\gamma) \SF(2\nu,e) \Big] \notag \\
&-\frac{64 \eta}{5 \alpha^{3} (1-e^{2})^{7/2}}\left( 1+\frac{73}{24}e^{2}+\frac{37}{96}e^{4} \right) , \label{eq:orbital_alpha} \\
\left\langle \frac{\dif e}{\dif\tau} \right\rangle=
&\zeta \alpha^{3/2} \sqrt{1-e^{2}} \Big[\Lambda_{2} \sin(\pi \nu + \gamma) \SF(\nu,e) \notag \\
&~~~~~~~~~~~~+ \Lambda_{4} \sin(2\pi \nu + 2\gamma) \SF(2\nu,e) \Big] \notag \\
&-\frac{304 \eta e}{15 \alpha^{4} (1-e^{2})^{5/2}}\left( 1+\frac{121}{304}e^{2} \right) \;, \label{eq:orbital_e} \\
\left\langle \frac{\dif \varphip}{\dif\tau} \right\rangle=
&-\zeta \alpha^{3/2} \frac{\sqrt{1-e^{2}}}{e} \Big[\Lambda_{2} \cos(\pi \nu + \gamma) \CF(\nu,e) \notag \\
&~~~~~~~~~~~~+ \Lambda_{4} \cos(2\pi \nu + 2\gamma) \CF(2\nu,e) \Big] \;, \label{eq:orbital_phip} \\
        \left\langle \frac{\dif \varphi}{\dif\tau} \right\rangle=& \alpha^{-3/2} \;, \label{eq:orbital_phi}
    \end{align}
\end{subequations}
where $\gamma = \Omega t_{\text{p}} + 2\Upsilon$.
Generally, it is possible to choose the initial time and coordinates in such a way that $\tp=0$ and $\varphi=\varphip=0$, as well as $\Upsilon=0$.

\begin{figure}[htbp]
    \includegraphics[width=\columnwidth]{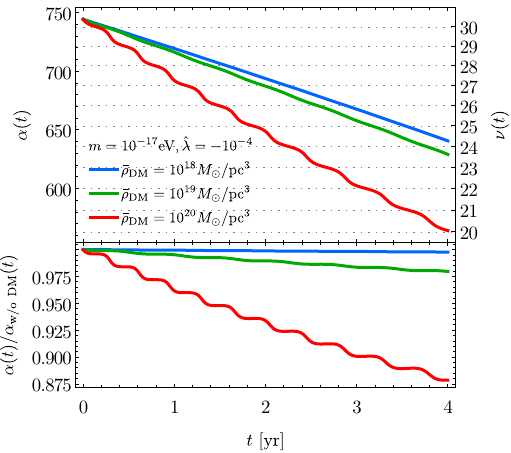}
    \caption{\textbf{Time evolution of the dimensionless semimajor axis $\alpha$ for an equal-mass binary system.} The binary system is characterized by a total mass $ M=10^{4}\Msun$, an initial orbital frequency $\omega_{0}=10^{-3}\Hz$, and an initial eccentricity $e_{0}=0.5$.
    Benchmark ULDM parameters include $m = 10^{-17}$eV and $\lh = -10^{-4}$ with average ULDM densities (blue, green, and red lines) given by \(\rhoDMb = \{10^{18}, 10^{19}, 10^{20} \} \Msun / \pc^3\) respectively.}
    \label{fig:PlotGrid_alpha}
\end{figure}

Fig.~\ref{fig:PlotGrid_alpha} shows how the dimensionless semimajor axis $\alpha= a/\Rx$ changes over time for a binary system with equal masses.
This system has a total mass $M=10^{4}\Msun$, an initial orbital frequency of $\omega_{0}=10^{-3}\Hz$, and an initial eccentricity of $e_{0}=0.5$.
We use the benchmark ULDM parameters: $m = 10^{-17}$eV, $\lh = -10^{-4}$, and $\rhoDMb = \{10^{18}, 10^{19}, 10^{20} \} \Msun / \pc^3$.
The lower subfigure displays the ratio of $\alpha$ to the vacuum scenario in the absence of ULDM cloud near the binary system.
As $\alpha$ decreases over time, causing $\omega$ to increase, the ratio $\nu$ consequently decreases, as indicated by the additional $y$-axis on the right side of the plot.
Notably, when the ULDM density is high, for instance $\rhoDMb = 10^{20} \Msun / \pc^3$, the plot reveals distinctive oscillatory features in $\alpha$, characterized by periodic dips occurring at specific intervals of $\nu$. This behavior highlights the dynamic interaction between the gravitational effects of the binary system and the surrounding ULDM environment.
The time evolution of other orbital elements, $e$, $\varphip$, and $\varphi$, is detailed in Appendix~\ref{sec:FDorbit}.

\section{Gravitational waves}
\label{sec:GWs}

The waveform of GWs originating from the inspiral of a binary system is described by~\cite{maggiore2008gravitational,Yunes:2009yz}
\begin{equation}
    \begin{aligned}\label{eq:h+}
        h_{+}(t)
 &=_{\ret} \frac{1}{1-e^{2}}\frac{4 (G \Mc)^{5/3} \omega^{2/3}}{\dL c^{4}} \bigg\{ \frac{1+\cos^{2}\iota}{2}\cos(2\varphi-2\beta) \\
 &+ \frac{e}{4}\sin^{2}\iota \Big[ \cos(\varphi-\varphip)+e \Big] \\
 &+\frac{e}{8} (1+\cos^{2}\iota) \Big[ 5\cos(\varphi-2\beta+\varphip) \\
 &+\cos(3\varphi-2\beta-\varphip) +2e\cos(2\beta-2\varphip) \Big] \bigg\} ,
    \end{aligned}
\end{equation}
\begin{equation}
    \begin{aligned}\label{eq:hx}
        h_{\times}(t)
 & =_{\ret} \frac{1}{1-e^{2}}\frac{4 (G \Mc)^{5/3} \omega^{2/3}}{\dL c^{4}} \bigg\{ \cos\iota \sin(2\varphi-2\beta) \\
 &+\frac{e}{4} \cos\iota \Big[ 5\sin(\varphi-2\beta+\varphip) \\
 &+\sin(3\varphi-2\beta-\varphip) -2e\sin(2\beta-2\varphip) \Big] \bigg\} ,
    \end{aligned}
\end{equation}
where `$=_{\ret}$' indicates that the right-hand side is computed at retarded time. The symbol $\Mc = \mu^{3/5} M^{2/5}$ represents the chirp mass, $\dL$ stands for the luminosity distance to source, $\iota$ denotes the angle between the orbital angular momentum axis of a binary and the direction to a detector, and $\beta$ represents the azimuthal component of the inclination angle.

\begin{figure}[htbp]
    \includegraphics[]{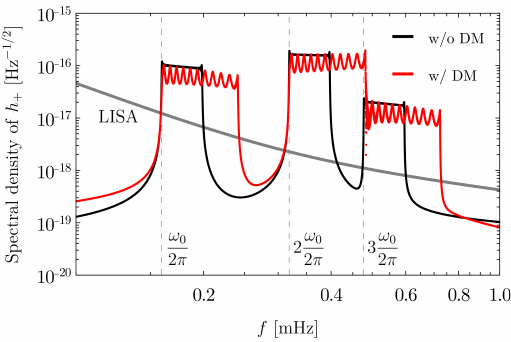}\\
    \includegraphics[]{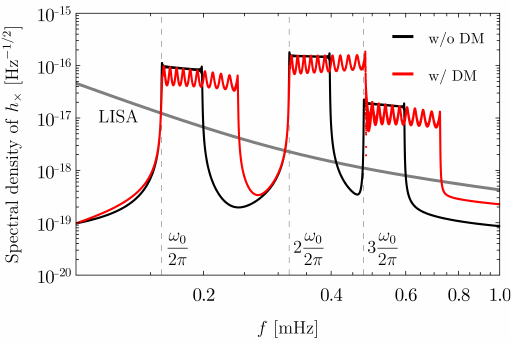}
    \caption{\textbf{The amplitude spectral density of GWs with (red line) and without (black line) ULDM clouds around the binary system.}
        The binary system has a total mass of $M = 10^{4}\Msun$, an initial orbital frequency of $\omega_{0} = 10^{-3}\Hz$, and an initial eccentricity of $e_{0} = 0.5$. It is located at a distance of $\dL = 0.1\Gpc$ and has orbital inclinations of $\iota = \pi/4$ and $\beta = \pi/4$.
        The benchmark ULDM parameters are same as those presented in Fig.~\ref{fig:PlotGrid_alpha}, except that the average ULDM density is fixed by $\rhoDMb = 10^{20} \Msun / \pc^3$. The gray line represents LISA's sensitivity curve~\cite{Robson:2018ifk}.
    }
    \label{fig:waveform}
\end{figure}

Taking the Fourier transformation of $h_{+}$ and $h_{\times}$, we can obtain the amplitude spectral density of GWs in the frequency domain as shown in Fig.~\ref{fig:waveform}. The red (black) line corresponds to the binary system with (without) ULDM clouds. Each broad peak corresponds to the $n$-th harmonic in the Fourier decomposition of the Keplerian motion. These peaks occur at $f_n = n\omega_{0}/2 \pi$ with $n \geq 1$.
In contrast to the vacuum scenario, the binary system surrounded by ULDM clouds exhibits a residual oscillatory pattern at each peak location. These non-trivial oscillations occur due to the oscillatory behavior of the binary orbital elements.
Since these oscillationg amplitudes exceeds LISA's sensitivity curve (gray line)~\cite{Robson:2018ifk}, in the future, identifying oscillatory patterns in GWs may indicate the existence of ULDM.

LISA operates in a heliocentric orbit and is composed of three spacecraft arranged in an equilateral triangle, with each spacecraft 2.5 million kilometers apart from the others. The constellation’s center of mass, known as the guiding center, moves in a circular orbit 1 AU away from the Sun and lags 20 degrees behind Earth. Using a polar coordinate system centered on the Sun, the strain of GWs at a detector is described by~\cite{Rubbo:2003ap}
\begin{equation}
    \begin{aligned}
        h(t) &= h_{+}(t-\Delta t) F_{+}(t-\Delta t, \theta_{\mathrm{s}}, \phi_{\mathrm{s}}, \chi) \\
             &+ h_{\times}(t-\Delta t) F_{\times}(t-\Delta t, \theta_{\mathrm{s}}, \phi_{\mathrm{s}}, \chi) ,
    \end{aligned}
\end{equation}
where $F_{+}$ and $F_{\times}$ represent the detector response functions, $\Delta t$ denotes the time delay between the arrival of GWs at the Sun and a detector, $\theta_{\mathrm{s}}$ and $\phi_{\mathrm{s}}$ are the latitude and longitude of source, and $\chi$ is the polarization angle of GWs.

When the signal-to-noise ratio (SNR) is high, the posterior probability distribution for the source parameters can be approximated as a multivariate Gaussian distribution, centered on the true values. The associated covariance matrix can be determined using the inverse of the Fisher information matrix. For a network of $N$ independent detectors, the Fisher matrix is expressed as
\begin{equation}\label{eq:Fisher}
    \Gamma_{ij} = \left( \frac{\partial \bm{d} (f)}{\partial \theta_{i}} , \frac{\partial \bm{d} (f)}{\partial \theta_{j}} \right)_{\bm{\theta}=\hat{\bm{\theta}}} \;,
\end{equation}
where $\bm{d}$ is given by
\begin{equation}
    \bm{d}(f)= \left[ \frac{\tilde{h}_{1}(f)}{\sqrt{S_{1}(f)}} , \frac{\tilde{h}_{2}(f)}{\sqrt{S_{2}(f)}} , \dots , \frac{\tilde{h}_{N}(f)}{\sqrt{S_{N}(f)}}\right]^{\mathrm{T}} \;,
\end{equation}
where $\bm{\theta}$ represents the vector of parameters with its true value denoted by $\hat{\bm{\theta}}$. In this context, $S_{i}(f)$ refers to the noise power spectral density of the $i$-th detector, and $\tilde{h}_{i}(f)$ is the Fourier transform of the signal in the time domain. The bracket operator $(A, B)$ for any two functions $A(t)$ and $B(t)$ is defined as
\begin{equation}
    (A,B) = 2 \int_{f_{\min}}^{f_{\max}} \dif f \left[ \tilde{A}(f)\tilde{B}^{*}(f) + \tilde{A}^{*}(f)\tilde{B}(f) \right] \;.
\end{equation}
The total SNR is given by $\sqrt{(\bm{d}, \bm{d})}$.

The root-mean-squared errors for the parameters can be derived from the inverse of the Fisher matrix
\begin{equation}
    \sigma_{\theta_{i}} = \sqrt{(\Gamma^{-1})_{ii}}\;.
\end{equation}
In our analysis, $\bm{\theta}$ consists of 14 parameters
\begin{equation}
    \bm{\theta}=\{ M, \eta, \omega_{0}, e_{0}, \varphi_{0}, \dL, \iota, \beta, \theta_{\mathrm{s}}, \phi_{\mathrm{s}}, \chi; \rhoDMb, m, \lh \} \;,
\end{equation}
where the first 11 parameters are related to the binary black holes and the last 3 parameters are related to ULDM.
The angles $\{ \iota, \beta, \theta_{\mathrm{s}}, \phi_{\mathrm{s}}, \chi \}$ are each set to $\pi/4$. We adjust the luminosity distance $d_{L}$ to vary SNR.

\begin{figure}[htbp]
    \includegraphics[width=0.48\textwidth]{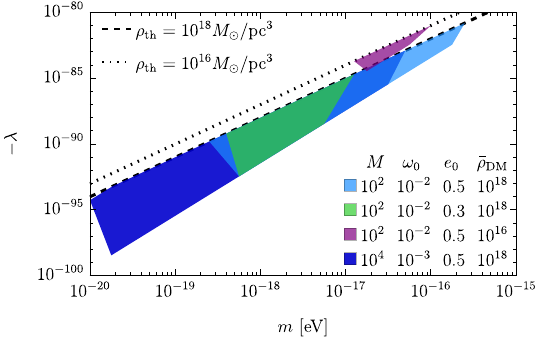}
    \caption{\textbf{The detectable regions in the parameter space of $\{ m, \lambda \}$ for binary systems under different conditions.} The reference parameters (highlighted in light blue) involve a binary system with $M=10^{2}\Msun$, an initial eccentricity $e_{0}=0.5$, and an average ULDM density $\rhoDMb = 10^{18} \Msun / \pc^3$.
        Additional scenarios shown include a system with a lower initial eccentricity $e_0 = 0.3$ (in green), another with a lower ULDM density $\rhoDMb = 10^{16} \Msun / \pc^3$ (in purple), and a binary system with an increased total mass of $M=10^{4}\Msun$ (in dark blue). The dotted and dashed lines mark the soliton stability thresholds at $10^{16} \Msun / \pc^3$ and $10^{18} \Msun / \pc^3$ respectively.
    }
    \label{fig:Main}
\end{figure}

Fig.~\ref{fig:Main} shows the detectable region within the parameter space of $\{ m, \lambda \}$ for an equal-mass binary system. This system has a total mass of $M=10^{2}\Msun$ and an initial orbital frequency of $\omega_{0}=10^{-2}\Hz$ (highlighted in light blue), referred to as the reference set of parameters.
This region corresponds to relative errors in mass and coupling, $\sigma_{m}/m$ and $\sigma_{\lh}/\lh$, to be less than 0.1, while also satisfying the stability requirement for solitons.
The SNR is assumed to be 100, the initial eccentricity is set at $e_{0}=0.5$, and the average ULDM density is fixed by $\rhoDMb = 10^{18} \Msun / \pc^3$.

To expand the detectable region to lower ULDM mass ranges, we consider a binary system with a larger total mass, $M=10^{4}\Msun$, which possesses a lower initial orbital frequency $\omega_0=10^{-3}\Hz$ (shown in dark blue).
With respect to the reference parameters, additional scenarios include a lower initial eccentricity $e_0 = 0.3$ (shown in dark green) and a smaller ULDM density $\rhoDMb = 10^{16} \Msun / \pc^3$ (shown in dark purple).
The parameter space that can be identified through resonating black holes can be further extended to higher ULDM mass by considering a binary system with a smaller total mass or a higher initial eccentricity. Investigating much lower absolute values of $\lambda$ requires a higher value of $\rhoDMb$.

\section{Conclusions}
\label{sec:dis}
We have investigated inspiralling binary black holes orbiting within a compact bosonic soliton and show that the binary’s chirping orbital frequency sweeps through harmonics of the boson’s Compton frequency, triggering a series of resonant crossings. This produces a distinctive modulation pattern in the GW signal.
We explored the new ULDM parameter space, and assessed how these parameters can impact the observable GW signals through LISA. This approach expands the range of ULDM searches, especially in situations where ULDM does not interact with the Standard Model particles.
This approach paves a way for broader applications.
It is applicable to ULDM with a positive self-coupling as well as to cases without self-interaction.
During inspiral, the component spins remain too small for supperradiance, so the resonant GW signature considered here provides an dominant effect.
After merger, the remnant's large spin could activates superradiance, making the two effects complementary at different stages of the binary's evolution.
Additionally, one can explore the multi-scalar field model, where two interfering ULDM fields may induce a beat phenomenon in GWs.
The implications of our research are significant for future experiments such as LISA, Taiji, Tianqin, DECIGO, and ET, which could detect these unique signals across wide frequency ranges.
As we conclude our discussion, it is important to acknowledge that further exploration is needed to understand the effects of dynamical friction on black holes moving through ULDM environment and halo feedback. We plan to address this topic in subsequent studies.

\textbf{Acknowledgments.}
We appreciate Sergey Sibiryakov, Clifford P. Burgess, Han Gil Choi, and Chang Sub Shin for the valuable discussions.
JHK is supported partly by the National Research Foundation of Korea (NRF) Grant NRF-2021R1C1C1005076, the BK-21 FOUR program through NRF, and the Institute of Information \& Communications Technology Planning \& Evaluation (IITP)-Information Technology Research Center (ITRC) Grant IITP-2025-RS-2024-00437284.
XYY is supported in part by the KIAS Individual Grant No. QP090702.

\appendix

\section{Stability Conditions for Soliton Configurations in ULDM}
\label{sec:stability}

The negative sign of the self-coupling $\lambda$ causes ULDM to interact attractively with each other. To form a stable soliton, a quantum pressure is required to counterbalance the attractive forces and gravity.
The mass of a stable soliton cannot exceed a specific threshold \cite{Chavanis:2011zi, Chavanis:2011zm, Levkov:2016rkk}
\begin{equation}
    \begin{aligned}
        M_{\rm th} \simeq& 10.2 \frac{m_{\rm pl}}{|\lambda|^{1/2}} \;,
    \end{aligned}
    \label{eq:Mth}
\end{equation}
where $m_{\rm pl} = \sqrt{1/G}$ denotes a Planck mass.
Beyond this threshold, the system is predicted to collapse.
The radius containing $99\%$ of the mass of the soliton is
\begin{equation}
    R_{99}
    \simeq 5.5 \Big( \frac{|b| }{G m^3} \Big)^{1/2}
    = 5.5 \Big( \frac{|\lambda| }{32\pi G m^4} \Big)^{1/2} \;,
    \label{eq:R99a}
\end{equation}
where $b = \lambda/(32 \pi m) $ denotes a scattering length.
The threshold density of the soliton is determined by
\begin{equation}
    \rho_{\rm th}
    \simeq 0.04 \frac{G m^4}{b^2}
    = 0.04 \frac{32^2 \pi^2 G m^6}{\lambda^2}
    \;.
    \label{eq:rhomax1}
\end{equation}

Our study focuses on a separation distance of approximately $\mathcal{O}(100) r_s$ between the binary black holes, given the Schwarzschild radius of the binary as $r_s = 2 GM /c^2 \sim 10^{-9} {\rm pc}$ for $M = 10^4 M_{\odot}$.
For the detectable parameter space of $m$ and $\lambda$, the soliton's size reaches approximately $R_{99} \sim 10^{-5} {\rm pc}$.
Therefore, the binary system is situated within the soliton's structure.

\section{Pressure Modulations in ULDM}
\label{sec:Pressure}

By utilizing Eq.(\ref{eq:rhoDM}), we are able to determine the amplitude, $\phi_{0}^{2} = \rhoDMb \Lambda/ m^2$,
where $\Lambda$ is given by
\begin{equation}
    \begin{aligned}
        \Lambda &= - \frac{16}{\lh} - \frac{8\big(-27 \lh^4 - 3 \sqrt{3} \lh^3 \sqrt{-64 + 27 \lh^2}~\big)^{1/3}}{3\lh^2} \\
                &- \frac{32}{ \big( -27 \lh^4 - 3\sqrt{3} \lh^3 \sqrt{-64 + 27 \lh^2}~\big)^{1/3} } \;,
    \end{aligned}
    \label{eq:Lambda}
\end{equation}
with a dimensionless parameter $\lh \equiv \lambda \rhoDMb/m^{4}$ defined in the interval $\lh \in [-\sqrt{64/27},0)$.

\begin{figure}[htbp]
    \centering
    \includegraphics[]{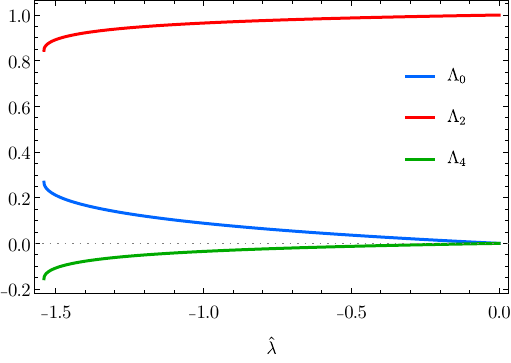}
    \caption{\textbf{Variation of dimensionless pressure components of ULDM as functions of $\lh$.} The frequency mode $\Lambda_{2}$ remains the dominant one throughout. }
    \label{fig:Lambda}
\end{figure}

The pressure can be expressed using these parameters as detailed in Eq.(\ref{eq:pDM}). Fig.~\ref{fig:Lambda} illustrates the behavior of the dimensionless parameters $\Lambda_{0}$, $\Lambda_{2}$, and $\Lambda_{4}$ as functions of $\lh$. As $\lh$ approaches zero, $\Lambda_2$ clearly dominates over $\Lambda_0$ and $\Lambda_4$, indicating that the oscillatory component of the pressure at frequency $2\omega_a$ (associated with $\Lambda_2$) is significantly larger than the other components. On the other hand, as $\lh$ approaches its lower limit, the magnitudes of $\Lambda_4$ and $\Lambda_0$ increase, but the contribution from $\Lambda_2$ remains comparatively dominant throughout, suggesting that its impact on the overall pressure modulation is large.

\begin{figure}[htbp]
    \centering
    \includegraphics[]{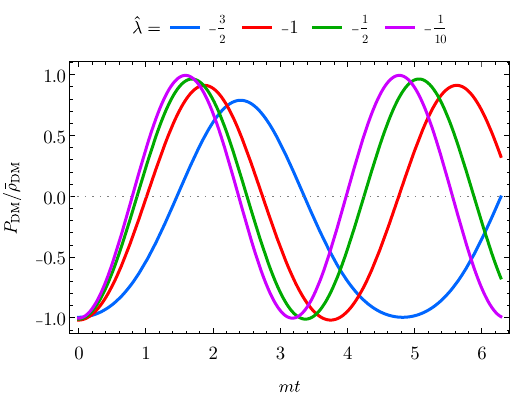}
    \caption{\textbf{Pressure modulations in ULDM, normalized by the average energy density, as a function of $m t$}.
        As $\lh$ approaches to its minimum value of $-\sqrt{64/27}$, the pressure undergoes significant changes.
    }
    \label{fig:Pressure}
\end{figure}

Fig.~\ref{fig:Pressure} shows the modulation of pressure, normalized by $\rhoDMb$, as a function of $mt$. As $\lh$ decreases to its minimum value of $-\sqrt{64/27}$, the ULDM angular frequency $\omega_{a}=m(1+\lh\Lambda/16)$ becomes smaller as illustrated by the blue line.
Our Fisher analysis, as explained in a later section, is highly sensitive to changes in pressure based on $\lh$. As the value of $\lh$ approaches it minimum, this sensitivity increases. Conversely, as $\lh$ approaches to zero, the variations in pressure exhibit less fluctuation. The observed trends suggest that the estimation error, $\sigma_{\lh}$, derived from Fisher analysis will be minimized as $\lh$ approaches its lower limit.
In other words, this parameter space enables more precise exploration, reducing the relative error and improving the accuracy of parameter measurements.

\section{Perturbed Einstein Equations}
\label{sec:Eins}

In this section, we solve the time-time and the space-space component of Einstein Eq.(\ref{eq:Ein00}) and Eq.(\ref{eq:Einij}) respectively, following a similar method as discussed in Ref~\cite{Khmelnitsky:2013lxt}.
The gravitational potential $\Psi$ consists of the time-independent components along with parts that oscillate at frequencies of $2\omega_a$ and $4\omega_a$. The proposed form for the solution of $\Psi$ is
\begin{equation}\label{eq:psi}
    \begin{aligned}
        \Psi (t, \bm{x}) = \bar{\Psi}(\bm{x}) &+ \tilde{\Psi}_2(\bm{x}) \cos(2 \omega_a t+2\Upsilon(\bm{x})) \\
                                              &+ \tilde{\Psi}_4(\bm{x}) \cos(4 \omega_a t+4\Upsilon(\bm{x})) \;,
    \end{aligned}
\end{equation}
where $\bar{\Psi}(\bm{x})$ denotes a time-independent component, and spatial gradients of $\tilde{\Psi}_2(\bm{x})$ and $\tilde{\Psi}_4(\bm{x})$ are assumed to be small. Likewise, the same applies for $\Phi(t,\bm{x})$. However, as this does not affect the final analysis, we will focus on $\Psi(t,\bm{x})$.

Taking time-independent parts from Eq.(\ref{eq:Ein00}) and taking a Fourier transformation, we find
\begin{equation}
    \bar{\Psi}(k) = - \frac{4 \pi G \rhoDMb}{k^2} \;.
    \label{eq:psib}
\end{equation}
The time-independent components of Eq.(\ref{eq:Einij}) gives
\begin{equation}
    \nabla^2 ( \bar{\Psi} - \bar{\Phi}) = 12 \pi G \rhoDMb \Lambda_{0} \;,
    \label{eq:eq:Einij2}
\end{equation}
which indicates that $\bar{\Psi} \neq \bar{\Phi}$.
Neglecting the spatial gradients, the time-dependent components of Eq.(\ref{eq:Einij}) gives
\begin{align}
    \tilde{\Psi}_2(\bm{x}) &= \frac{\pi G \rhoDMb \Lambda_2}{\omega^2_a} \;, \\
    \tilde{\Psi}_4(\bm{x}) &= \frac{\pi G \rhoDMb \Lambda_4}{4\omega^2_a} \;,
    \label{eq:eq:Einij3}
\end{align}
Differentiating $\Psi$ with respect to time twice, we have
\begin{equation}
    \ddot{\Psi} = - 4 \pi G \rhoDMb \Big[ \Lambda_2 \cos(2\omega_a t+2\Upsilon) + \Lambda_4 \cos(4\omega_a t+4\Upsilon) \Big] \;.
    \label{eq:psidd}
\end{equation}
This metric perturbation generates the additional force between the black holes, $\FDM=\ddot{\Psi}r$, as described in Eq.(\ref{eq:ddr}).

\section{Averaged Orbital Dynamics}
\label{sec:FDorbit}

To solve the orbital evolution including the contribution from the emission of GWs, we compute the averages of these quantities over an orbital period, $T=2\pi/\omega$, as follows
\begin{equation}
    \langle\cdots\rangle \equiv \int_{0}^{T} \frac{dt}{T}(\cdots) =\int_{0}^{2\pi} \frac{d\xi}{2\pi}(\cdots) \;,
\end{equation}
where $\xi \equiv \omega \ttld$ and $\ttld\equiv t -\tp$.
By using the Kepler's equation~\cite{poisson2014gravity}
\begin{equation}
    u-e\sin u = \omega(t-\tp) \;,
\end{equation}
where $u$ denotes an eccentric anomaly and $\tp$ is a time of pericenter passage, we consider the Fourier decomposition of Keplerian functions~\cite{watson1995treatise,Blas:2019hxz}
\begin{align}
    \frac{r}{a}\sin\varphit =& \frac{2\sqrt{1-e^{2}}}{e}\sum_{n=1}^{\infty}\frac{J_{n}(ne)}{n}\sin(n\omega\ttld) \;, \\
    \frac{r}{a}\cos\varphit =& -\frac{3e}{2} + 2\sum_{n=1}^{\infty}\frac{J_{n}'(ne)}{n}\cos(n\omega\ttld) \;,
\end{align}
where $\varphit \equiv \varphi - \varphip$ denotes a true anomaly,
$J_{n}(z)$ is the Bessel function, and $J_{n}'(z)$ is its derivative with respect to $z$.
The Fourier decomposition is useful because it allows the complicted motion of binary system to be approximated by Keplerian orbits.

Incorporating the oscillating part $\cos(\Omega\ttld+\gamma)$ due to the ULDM, where $\gamma = \Omega t_{\text{p}} + 2\Upsilon$, the time-averaged quantities can be computed by
\begin{equation}
    \begin{aligned}
&\left\langle \frac{r}{a}\sin\varphit \cos(\Omega\ttld+\gamma)\right\rangle \\
&= \int_{0}^{2\pi} \frac{d\xi}{2\pi}\left[\frac{2\sqrt{1-e^{2}}}{e}\sum_{n=1}^{\infty}\frac{J_{n}(ne)}{n}\sin(n\xi)\right] \cos(\nu\xi+\gamma)\\
&=\sin(\pi\nu+\gamma)\SF(\nu,e) \;,
    \end{aligned}
    \label{eq:AvSF}
\end{equation}
and
\begin{equation}
    \begin{aligned}
&\left\langle \frac{r}{a}\cos\varphit \cos(\Omega\ttld+\gamma)\right\rangle \\
&= \int_{0}^{2\pi} \frac{d\xi}{2\pi}\left[-\frac{3e}{2} + 2\sum_{n=1}^{\infty}\frac{J_{n}'(ne)}{n}\cos(n\xi)\right] \cos(\nu\xi+\gamma)\\
&=\cos(\pi\nu+\gamma)\CF(\nu,e) \;,
    \end{aligned}
    \label{eq:AvCF}
\end{equation}
where special functions, $\SF(\nu,e)$ and $\CF(\nu,e)$, are defined by
\begin{equation}
    \begin{aligned}
&\SF(\nu,e)=\frac{2\sqrt{1-e^{2}}\sin(\pi\nu)}{\pi e}\sum_{n=1}^{\infty}\frac{J_{n}(ne)}{n^{2}-\nu^{2}}\\
&=\left\{
    \begin{aligned}
&0, ~\nu=0\\
&\frac{(-1)^{\nu-1}\sqrt{1-e^{2}}}{2\nu} \left[ J_{\nu-1}(\nu e) + J_{\nu+1}(\nu e) \right], ~\nu=1,2,\cdots\\
&\frac{\sqrt{1-e^{2}}}{2\pi\sin(\pi\nu)}\times \\&\int_{0}^{2\pi}\dif u (1-e\cos u)\sin u \cos[\nu (u-e\sin u)], ~\nu \not\in \mathbb{Z}
    \end{aligned}
\right.
    \end{aligned}
    \label{eq:SF}
\end{equation}
\begin{equation}
    \begin{aligned}
&\CF(\nu,e)=-\frac{\sin(\pi\nu)}{\pi \nu} \left[ \frac{3e}{2}+ 2\sum_{n=1}^{\infty}\frac{J_{n}'(ne) \nu^{2}}{(n^{2}-\nu^{2})n} \right] \\
&=\left\{
    \begin{aligned}
&-\frac{3e}{2}, ~\nu=0\\
&\frac{(-1)^{\nu}}{2\nu} \left[ J_{\nu-1}(\nu e) - J_{\nu+1}(\nu e) \right], ~\nu=1,2,\cdots\\
&\frac{1}{2\pi\sin(\pi\nu)}\times \\&\int_{0}^{2\pi}\dif u (1-e\cos u)(\cos u -e) \sin[\nu (u-e\sin u)], ~\nu \not\in \mathbb{Z}
    \end{aligned}
\right.
    \end{aligned}
    \label{eq:CF}
\end{equation}
Note that the orbital expansions in terms of Bessel functions can be expressed as piecewise functions in an integral form. Therefore, we employed these piecewise functions in our calculations, incorporating all terms in the expansions.

\begin{figure}[htbp]
    \centering
    \includegraphics[]{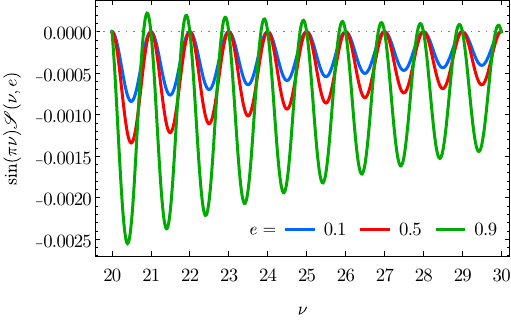}\\
    \vspace{5mm}
    \includegraphics[]{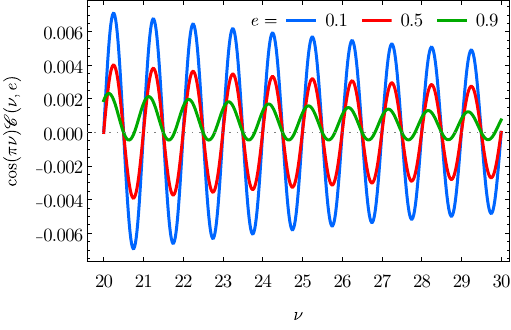}
    \caption{\textbf{Special functions that influence the dynamical evolution of the orbital elements in a binary black hole system with oscillating ULDM.} These functions are defined in Eq.(\ref{eq:AvSF}) (top) and Eq.(\ref{eq:AvCF}) (bottom) as a function of $\nu$ for various values of $e = 0.1, 0.5, 0.9$. As $\nu$ decreases over time, both functions exhibit periodic oscillations. Notably, peaks occur periodically within intervals of $\nu$.
    }
    \label{fig:SFCF}
\end{figure}

Fig.~\ref{fig:SFCF} shows the evolution of Eq.(\ref{eq:AvSF}) and Eq.(\ref{eq:AvCF}) as a function of $\nu$ for various values of eccentricities, $e = 0.1, 0.5, 0.9$. As the separation between the binary components decreases over time, causing $\omega$ to increase, the ratio $\nu$ consequently diminishes as the binary nears coalescence. As $\nu$ decreases over time, both functions exhibit periodic oscillations. Notably, peaks occur periodically within intervals of $\nu$.

\begin{figure}[htbp]
    \centering
    \includegraphics[]{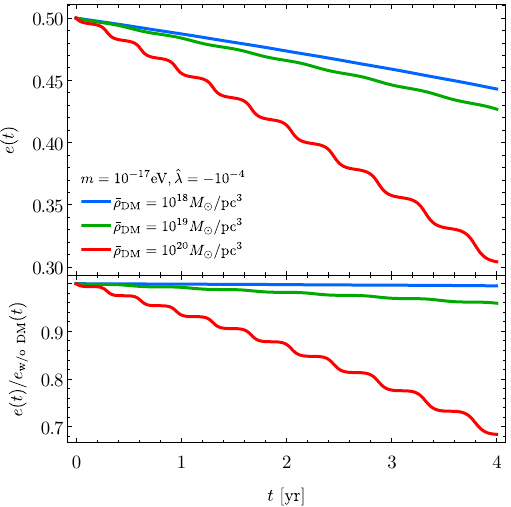}\\
    \vspace{5mm}
    \includegraphics[]{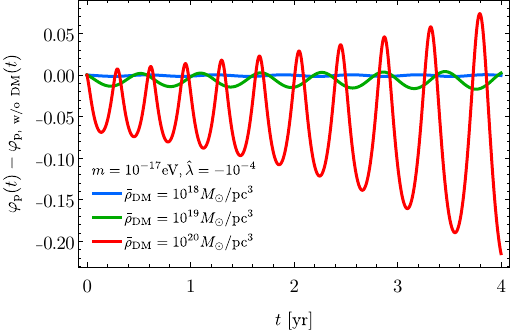}
    \caption{\textbf{Time evolution of the eccentricity $e$ and the longitude of the pericenter $\varphip$ for a binary system when influenced by oscillating ULDM.} We consider the binary system characterized by a total mass $M=10^{4}\Msun$, an initial orbital frequency $\omega_{0}=10^{-3}\Hz$, and an initial eccentricity $e_{0}=0.5$. Benchmark ULDM parameters are $m = 10^{-17}$eV and $\lh = -10^{-4}$.
    }
    \label{fig:Evolution_e}
\end{figure}

In Eq.(\ref{eq:orbital_alpha}), the function in Eq.(\ref{eq:AvSF}) plays a crucial role in influencing the dynamics of the dimensionless semi-major axis $\alpha$ and the eccentricity $e$ due to interactions with ULDM as shown in Fig.~\ref{fig:PlotGrid_alpha} and Fig.~\ref{fig:Evolution_e} (top panel) respectively. As time progresses and $\nu$ decreases, the Eq.(\ref{eq:AvSF}) oscillates and mainly becomes a negative value. This contributes to a resonant force that acts to reduce the value of $\alpha$ and $e$, at the points where $\nu$ roughly reaches integer values.

In Eq.(\ref{eq:orbital_phip}), the function in Eq.(\ref{eq:AvCF}) is important for determining the variations in the longitude of the pericenter $\varphip$, due to the resonant interactions with ULDM, as illustrated in Fig.~\ref{fig:Evolution_e} (bottom panel). As $\nu$ decreases over time, the function Eq.(\ref{eq:AvCF}) exhibits oscillatory behavior around zero. This oscillation is responsible for the similarly oscillatory pattern observed in $\varphip$, as it fluctuates around zero.

\begin{figure}[htbp]
    \centering
    \includegraphics[]{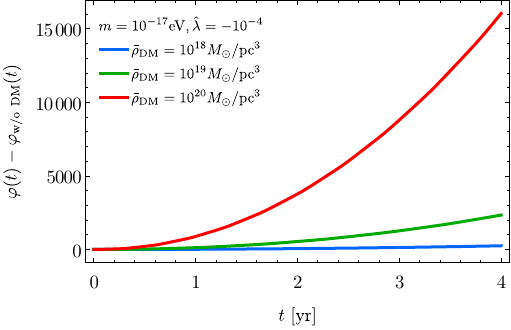}
    \caption{\textbf{The phase difference in the orbital angle of a binary system when influenced by oscillating ULDM compared to a vacuum scenario.} The binary system and benchmark ULDM parameters are same as those presented in Fig.~\ref{fig:Evolution_e}.
    }
    \label{fig:Evolution_phi}
\end{figure}

Finally, Fig.~\ref{fig:Evolution_phi} illustrates the phase difference in the orbital angle of a binary system when influenced by ULDM compared to a vacuum scenario. This phase difference shows how the dynamics of the binary system, affected by resonant forces from ULDM, result in a faster merger compared to vacuum scenarios.

\section{Axion Miniclusters and Solitons}
\label{sec:Mini}

Axion-like particle (ALP) is an attractive candidate of ULDM.
The potential of ALP is typically characterized by a periodic form $V(\phi) = m^2 f_a^2 \big( 1 - \cos (\phi/f_a) \big)$ where $f_a$ denotes its decay constant.
For small field values, the expression simplifies to approximately $V(\phi) \approx m^2\phi^2/2 - m^2 \phi^4/(4! f^2_a) $. Consequently, a negative quartic self-coupling constant arises, denoted as $\lambda = -m^2/f^2_a$.
As an example, we consider axion miniclusters (sometimes referred to as axion minihalos), which are dense objects formed by the collapse of large-amplitude isocurvature perturbations seeded during the phase transition that produces axion DM (for brevity, we use ``axion'' interchangeably with ``axionlike particle'')~\cite{Hogan:1988mp, Kolb:1993zz, Kolb:1993hw, Kolb:1994fi, Kolb:1995bu, Zurek:2006sy, Hardy:2016mns, Enander:2017ogx}.
These miniclusters can reach present-day densities exceeding $10^{14} \Msun/\pc^3$ through perturbations on scales corresponding to causally disconnected regions at $T \sim 1\GeV$~\cite{Kolb:1993zz, Kolb:1993hw}.
Their cores can host overdense solitons (also called axion stars) due to Bose–Einstein condensation driven by field relaxation~\cite{Levkov:2018kau, Chen:2020cef, Kirkpatrick:2020fwd}.
We focus on the densities of these soliton cores inside axion miniclusters.

Following Refs.~\cite{Fairbairn:2017dmf, Fairbairn:2017sil, Maseizik:2024qly}, the characteristic mass $M_0$ of a minicluster is determined by the total mass of axion dark matter within the horizon at the oscillation temperature $T_{\osc}$, defined by $3H(T_{\osc})=m_{T}(T_{\osc})$.
This mass depends on the low-temperature value of the axion mass $m$, the axion decay constant $f_a$, and the index $p$ that specifies the temperature evolution $m_{T}(T)=(T/\sqrt{m f_a})^{-p} m$.
The numerical value of $p$ is determined by the ultraviolet properties of the hidden gauge sector, in particular, the number of light degrees of freedom and the $\beta$-function coefficient. In what follows, we remain agnostic about the hidden sector dynamics and treat $p$ as a free parameter, scanning the range $0\le p\le 10$.

Inside a host galaxy such as the Milky Way, the minicluster mass function, denoting the comoving number
density of miniclusters of mass $M$ per logarithmic mass interval, can be parametrized by a power-law
\begin{equation}
    \frac{\dif n}{\dif \ln M} = C \left( \frac{M}{M_{\min}}\right)^{-1/2} ,
    \label{eq:MCMF}
\end{equation}
where $C$ is a normalization constant, and at redshift $z = 0$, the mass $M$ ranges approximately from $M_{\min} \approx M_0/25$ to about $M_{\max} \approx 4.9\times 10^6 M_0$~\cite{Fairbairn:2017sil}.
For the Milky Way, the normalization constant is determined by $M_{\MW} f_{\mc} = V_{\MW} \int_{M_{\min}}^{M_{\max}} \dif M \frac{\dif n}{\dif \ln M}$, where $M_{\MW}=1.43\times 10^{12}\Msun$ is the DM halo mass, $f_{\mc} \simeq 0.75$ encodes the fraction of dark matter contained in miniclusters, and $V_{\MW}$ denotes the Milky Way volume.

The mass of soliton cores inside miniclusters is given by the core-halo relation~\cite{Schive:2014hza,Eggemeier:2019jsu}
\begin{equation}
    M_* = M_{\hmin}\left( \frac{M}{M_{\hmin}} \right)^{1/3} ,
    \label{eq:halo_core}
\end{equation}
where $M_{\hmin}=2.36\times 10^{-16}\Msun \left( m/50\mueV \right)^{-3/2}$ denotes the lightest minicluster mass, at which a composite core-halo system can form.
The soliton core masses range from $M_*(M_{\min})$ to $M_{*,\max}=\min\{M_*(M_{\max}), M_{\rm th}, M_{\mathrm{relax}}\}$, where $M_{\rm th}$ denotes the instability limit defined in Eq.(\ref{eq:Mth}), and $M_{\mathrm{relax}} \simeq 10^{-13}\Msun \sqrt{\delta^3(1+\delta) \tau_{\mathrm{U}}/(1\Gyr)} \left(m/26\mueV\right)^{-3/2}$ is obtained from the relaxation time, assuming $\tau_{\mathrm{U}} \simeq 13.8\Gyr$ and $\delta \simeq 1$~\cite{Levkov:2018kau, Bar:2019pnz}.
The soliton core's central density, $\rho_*$, and the characteristic radius, $R_*$, are given by the mass-radius relation~\cite{Chavanis:2011zi}
\begin{eqnarray}
    \rho_* &=& \frac{M_*}{\pi^{3/2} R_*^3} , \\
    R_* &=& \sqrt{\frac{9\pi}{8}}\frac{1}{G M_* m^2} \left( 1+ \sqrt{1-\frac{G M_*^2 m^2}{12\pi^2 f_a^2}} \right) .
\end{eqnarray}

By combining the minicluster mass function in Eq.(\ref{eq:MCMF}) with the core-halo relation in Eq.(\ref{eq:halo_core}), and using the identity $\dif n = \dif n_*$, we can obtain the following expression for the mass distribution of soliton cores~\cite{Maseizik:2024qly,Maseizik:2024qya}
\begin{equation}
    \frac{\dif n_*}{\dif \ln M_*} = 3C \left( \frac{M_*}{M_{\hmin}}\right)^{-3/2} \left( \frac{M_{\hmin}}{M_{\min}}\right)^{-1/2} .
\end{equation}

Note that post-inflation Peccei--Quinn (PQ) breaking scenario enforces $f_a \lesssim H_I/(2\pi) \simeq 8 \times 10^{12}~{\rm GeV}$~\cite{Fairbairn:2017sil,BICEP2:2015xme} where $H_I$ denotes the Hubble parameter during inflation.
After inflation, however, if the reheating temperature is sufficiently high, the PQ symmetry can be restored.
In this case, when the symmetry subsequently breaks and the non-perturbative axion potential is generated, the misalignment angle $\theta = \langle a \rangle/f_a $ once again develops inhomogeneous fluctuations.
Thus, even if $f_a$ exceeds the inflationary bound, spatial variations in $\theta$ may still arise through this non-inflationary mechanism.
The energy density during inflation is given by $\rho_I = H_I^2 m_{\rm pl}^2$, which can be expressed in terms of a temperature scale via $\rho_I \sim T^4$. Consequently, the reheating temperature satisfies $T_R \lesssim \sqrt{H_I m_{\rm pl}} \sim 10^{16}~\mathrm{GeV}$.
If reheating occurs at such high scales, the PQ symmetry can be restored, allowing the axion minicluster scenario to remain viable.
Therefore, the condition $f_a \lesssim 8 \times 10^{12}~{\rm GeV}$ should be regarded not as a strict upper bound but as the most conservative criterion.
In practice, through non-inflationary mechanisms, axion miniclusters may still form for values as large as $f_a \sim 10^{16}~\mathrm{GeV}$.

\begin{figure}[htbp]
    \centering
    \includegraphics[width=0.48\textwidth]{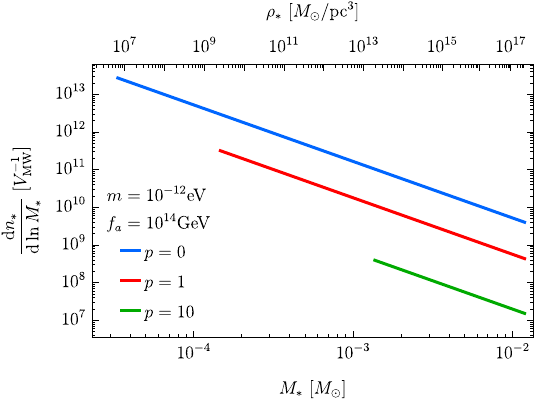}
    \caption{\textbf{Mass distribution of soliton cores per Milky Way volume for $m=10^{-12}\eV$ and $f_a=10^{14}\GeV$.} Colored lines indicate masses of soliton cores at different $p=0, 1, 10$.
    }
    \label{fig:dnxdlnMx}
\end{figure}

For illustration, Fig.~\ref{fig:dnxdlnMx} shows the mass distribution of soliton cores per Milky Way volume for $m=10^{-12}\eV$ and $f_a=10^{14}\GeV$.
The colored lines denote the masses of soliton cores for various values of the temperature evolution indices, $p=0, 1, 10$. The upper horizontal axis shows the corresponding central density of the solitons.
For $p=0$, roughly $10^9$ compact solitons, each with densities reaching up to about $5\times 10^{17}\Msun/\pc^3$ can exist per Milky Way volume. For higher $p$ values, even denser solitons can form, but their number density decreases significantly.

\begin{figure*}[t!]
    \centering
    \includegraphics[width=.32\textwidth]{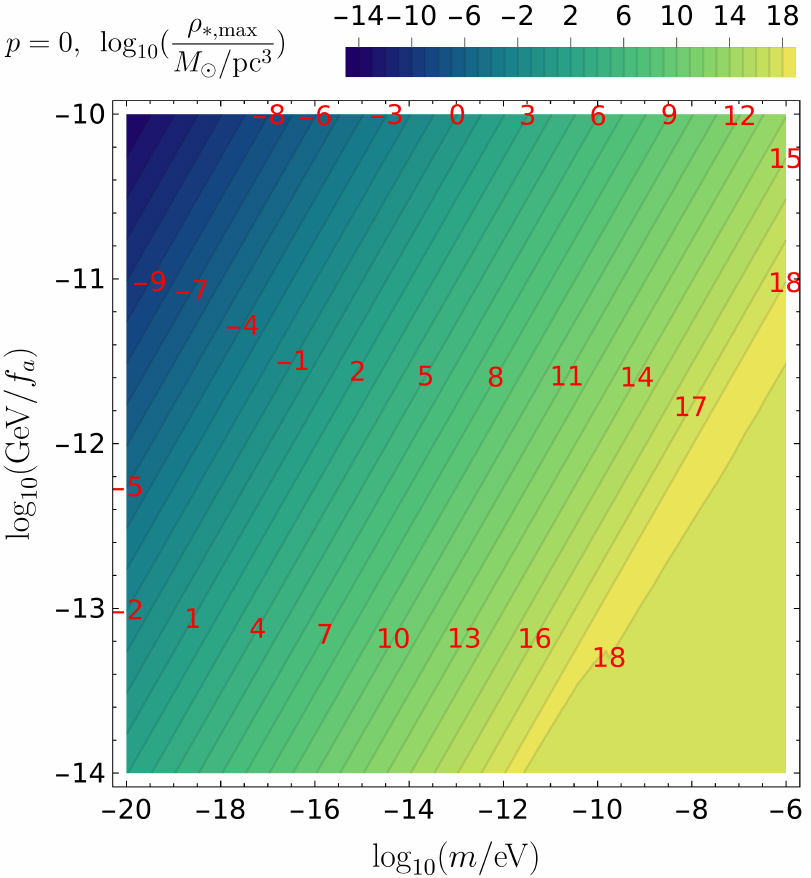}
    \includegraphics[width=.32\textwidth]{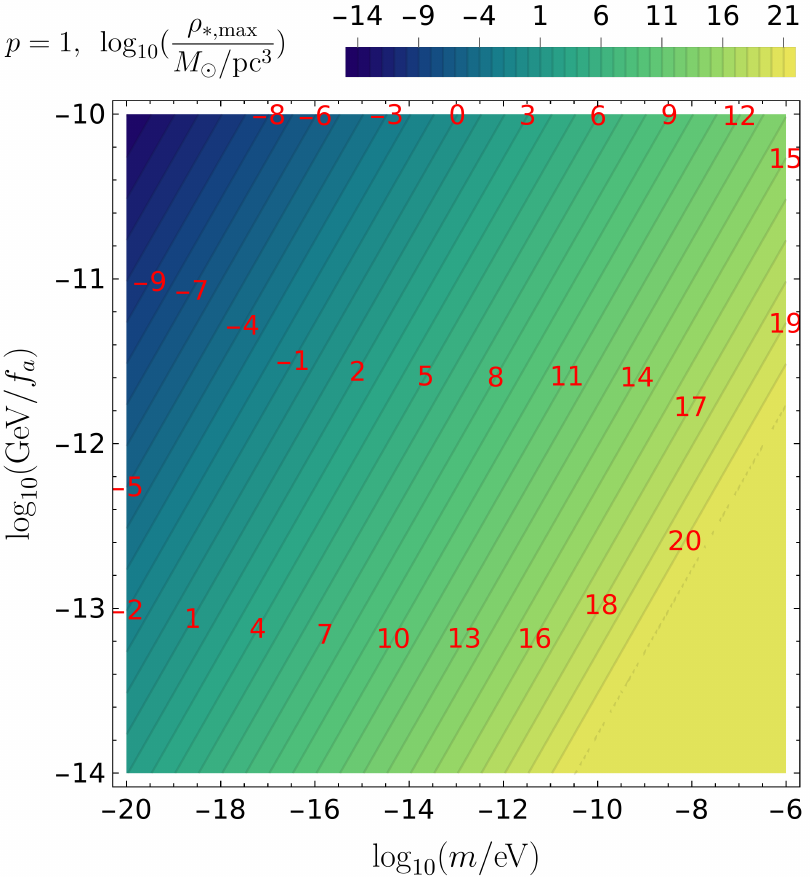}
    \includegraphics[width=.32\textwidth]{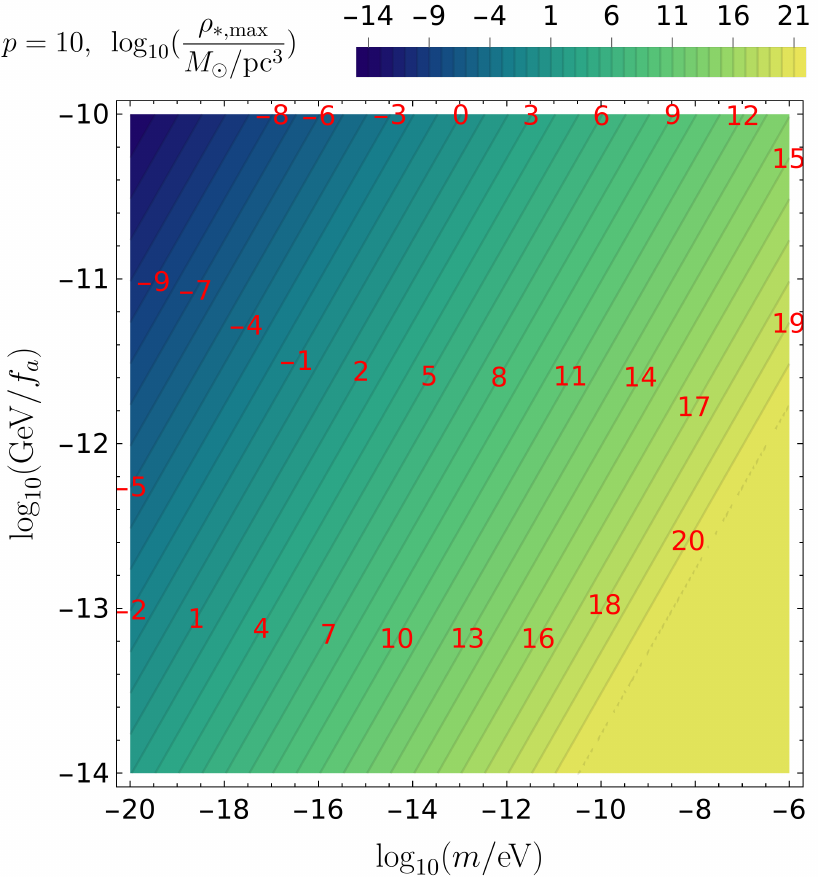}
    \caption{\textbf{Maximum achievable central densities of soliton cores as a function of $m$ and $f_a$.} The left panel corresponds to $p=0$, the middle to $p=1$, and the right to $p=10$.
    }
    \label{fig:rhoxmax}
\end{figure*}

In Fig.~\ref{fig:rhoxmax}, we present the maximum achievable central density of soliton cores in the axion minicluster scenario as a function of $m$ and $f_a$. The left panel corresponds to $p=0$, the middle to $p=1$, and the right to $p=10$. A high density soliton is achievable in the relatively heavy mass range $m \gtrsim 10^{-12}~{\rm eV}$.
However, at such masses, the resulting resonant GW frequency lies far above the LISA band. That high-frequency of $\mathcal{O}({\rm kHz})$ window might be accessible to detectors such as aLIGO~\cite{LIGOScientific:2014pky} or ET~\cite{Punturo:2010zz}.
Because LISA itself could not test that range of parameter space, we adopt in the main manuscript a more agnostic framework in which the boson mass $m$ and the quartic self-coupling $\lambda$ are treated as independent parameters, and we focus on how a generic high-density soliton modifies the gravitational waveform of a binary black-hole system.

Turning to the chance of finding a soliton that contains a binary black hole system, the probability depends on whether the black holes are astrophysical or primordial in origin.
We begin by estimating the merger rate for binaries of astrophysical-origin black holes residing inside high density solitons.
We denote $n_{\star}\equiv N_{\rm sol}/V_{\rm MW}$ the comoving number density of high-density solitons, normalized to the Milky-Way-equivalent volume. A fiducial value is $n_{\star}\sim 10^{9}\,V^{-1}_{\rm MW}$ for solitons with central densities of order $\sim 5 \times 10^{17}\Msun/\pc^3$ (see Fig.~\ref{fig:dnxdlnMx}).
We denote $p_{\rm BBH}$ the probability of two binary black holes reside inside a soliton.
Ref.~\cite{Lee:2025qbu} quotes an intrinsic intermediate mass black hole (IMBH) binary merger rate of
$\mathcal R_{\rm IMBH}\simeq 10\text{–}30\;{\rm Gpc}^{-3}\,{\rm yr}^{-1}$
within the LISA horizon ($z\lesssim0.5$).
Hence the expected rate of mergers inside solitons is
$\mathcal R_{\star{\rm BBH}} \simeq \mathcal R_{\rm IMBH}\,p_{\rm BBH} \,n_{\star} V_{\rm tot} $ where $V_{\rm tot} = {\rm 1Gpc}^{3}$.
For example, taking $p_{\rm BBH}\!=\!10^{-20}$ yields roughly $10$ events per year in the $1\;{\rm Gpc}^{3}$ comoving volume.
Note that for a SNR of $\mathcal{O}(100)$, the binary systems we consider are estimated to lie at a luminosity distance $d_L$ of about $\sim 1\Gpc$. In future GW surveys, the detection range is expected to expand, which will enhance the ability to observe these signal events.

The probability that a single soliton simultaneously hosts two distinct binary black holes may be significantly small. A more realistic configuration is that one black hole is embedded in a soliton. We denote by $f_{\rm BH}$ the fraction of all black holes that resides inside a soliton. A secondary astrophysical compact object with a small mass ratio $q$ can then be captured by this system and undergo an inspiral.
Estimating $p_{\rm BBH}$ and $f_{\rm BH}$ will require a dedicated numerical study. One would have to tract the high-density solitons and follow their subsequent interaction with black holes to obtain dynamical capture rate. Because such simulations are computationally demanding, we plan to determine $p_{\rm BBH}$ and $f_{\rm BH}$ in future work.

If primordial black holes (PBHs) exist, they serve as natural local overdensities within the DM distribution and thus act as seeds for further structure formation. In scenarios where axions constitute the dominant DM component, axion miniclusters are expected to grow around these PBHs \cite{Hertzberg:2020hsz,Yin:2024xov}, potentially leading to the formation of axion solitons in their vicinity. When these axion solitons that host PBHs merge, they can give rise to binary black holes.
Although a comprehensive feasibility study of this process is beyond the scope of the current work, preliminary estimates, which rely solely on the PBH merging rate, suggest that the event rate is around $0.01\Gpc^{-3}\yr^{-1}$ for PBHs with mass $100\Msun$ and abundance $f_{\mathrm{PBH}}=10^{-4}$~\cite{Ali-Haimoud:2017rtz, Raidal:2018bbj, Carr:2020gox}.

\bibliography{citeLib}

\end{document}